\documentstyle[preprint,aps,prb]{revtex}
\draft
\preprint{CBPF-NF-0411/96, PG-115/96}
\begin{document}
\title{Analytic calculations of trial wave functions\\  
of the fractional quantum Hall effect on the sphere}
\author{Carmem Lucia de Souza Batista \cite{byline0}}
\address{Centro Brasileiro de Pesquisas Fisica-CBPF \\
Rua Xavier Sigaud 150, 22290-180 Urca, RJ, Brazil}
\author{Dingping Li \cite{byline}}
\address{Dipartimento di Fisica and Sezione I.N.F.N.,
Universit\'a di Perugia \\
Via A. Pascoli,  I-06100 Perugia, Italy}
\date{\today}
\maketitle
\begin{abstract}
We present a framework for the analytic calculations of 
the hierarchical wave functions and the 
composite fermion  wave functions 
in the fractional quantum Hall effect
on the sphere  by using projective coordinates.
Then we calculate the overlaps
between these two wave functions at various 
fillings and small numbers of electrons.
We find that the overlaps are all most equal to one. 
This gives a further evidence that 
two theories of the  fractional quantum Hall effect,
the hierarchical theory and the composite fermion
theory,  are physically equivalent. 
\end{abstract}
\pacs{PACS: 73.40.Hm, 73.20.Dx,03.65.-w,03.80.+r}

\narrowtext

\section{Introduction}

The fractional quantum  Hall effect (FQHE) 
at the Landau-level (LL) filling fraction $\nu =1/m$
with $m$ an old integer is very well described by Laughlin's
theory.\cite{Laughlin2,PrangeGirvin} 
The Laughlin wave function   is  a very good approximation  
of the exact ground state of the  
quantum  Hall effect (QHE) at $\nu =1/m$. 
However for the FQHE at $\nu \not= 1/m$,
there exist {\it  two} well-known theories 
(notice that we will only consider
the case that the electron spins are polarized  in this paper).
One   is  the hierarchical theory.  The states at $\nu \not= 1/m$
are formed  due to the condensation 
of the anyonic quasiparticles of Laughlin states.
\cite{Haldane605,Halperin1,frast,other,Read,wenblok,greiter,jian}
The trial wave functions constructed from this theory 
are called as  the hierarchical wave functions.   
Another theory is based on  
the composite fermion (CF) approach proposed by
Jain,\cite{Jain3} where the FQHE is due to the integer QHE  of
the composite fermions (CFs) (electrons bounded with an 
even magnetic flux quanta).  The trial 
wave functions constructed from the CF theory
are called as the CF wave functions (or Jain's wave functions)  
The overlaps  of the exact  states  
with the  hierarchical  wave functions
and the CF wave functions are both excellent.
It has also been shown that two theories predict the same topological
excitations at the same  $\nu$ .\cite{Read,wenblok,wenr} 
The two theories  must be physically equivalent if they both
describe   correctly  the physics of the FQHE.
Thus it would be very interesting to study the difference
and equivalence of the two theories.

In this paper, we present a framework for the analytic calculations of 
the two wave functions on the sphere 
by using projective (or stereographic)
coordinates on the sphere. There are several advantages of using   
spherical geometry. As it is a compact surface, there will be no  edge 
state to be worried if we are only interested in the bulk state.
Also the system has rotational invariance  symmetries. 
On the torus, though  the system
has  translational invariance and no boundaries, 
the hierarchical wave functions are  very difficult to
calculate and  quite complicated due to
its nontrivial topology,\cite{torus} and we do not even know how 
to construct the CF  wave functions with the correct center 
coordinate degeneracy on a torus.  

Because the states considered in the FQHE 
are  restricted to the lowest 
Landau level (LLL), the wave functions are
only dependent on holomorphic coordinates (polynomials 
of the  holomorphic coordinates) on the sphere.
Therefore it is possible to use only holomorphic coordinates to do all 
calculations. To compare the two types of  hierarchical wave functions
is the same to compare the two polynomials of holomorphic coordinates
on the sphere.  We note that our ultimate 
goal is to expand those wave functions in polynomials and 
calculate the overlaps of two wave functions or physical quantities 
(for example, the  density-density  correlations)
at an {\it arbitrary number}  of electrons by the method  
(Jack polynomials method) used 
in studying the Calogero model.\cite{ha}   
We do not know how to do it at the moment, and
further progresses  on it will enhance 
our understandings of the theories of the FQHE.

We organize the paper as follows; 
first we review the Landau level problem on
the sphere. A self-contained derivation of eigenstates
of an electron on a sphere with a monopole field by using a simple
geometric argument and projective coordinates is given in the appendix.
Then we  show  how to classify the many-body eigenstates 
of the angular momentum in the LLL.  
We then construct the wave functions based 
the theory of the hierarchical states and the theory
based on the CF picture.
The wave functions constructed in this paper are 
easy to handle in the practical calculation.
Finally  we calculate the overlaps of 
the hierarchical wave functions  and the CF  wave functions
at various fillings $\nu$ and some small numbers of electrons.

\section{Quantum Mechanics On The Sphere}

  The electrons are constrained to move on the surface of a sphere of
radius $R$ having a magnetic monopole 
in its center.  The total magnetic 
flux $4\pi R^2B$ must be an integer multiple $\phi=2S$ of 
the magnetic flux quantum $\phi_0=2 \hbar \pi c/e$
according to the Dirac quantization condition.
Therefore, the sphere radius $R$ is equal to $S^{1/2}l_0$,  
where $l_0=(\case{\hbar c}{eB})^{1/2}$
is the magnetic length. The eigenstates of an electron 
are given by  monopole
spherical  harmonics.\cite{Haldane605,wuyang}
First, we  briefly review the old method 
to derive the wave functions of the Landau levels (LLs), 
then rederive them by using Algebraic Geometry. 

For simplicity, we take units $\hbar$ and $c$ equal to one
in the following formulas.
The Hamiltonian  of a single electron of mass $m_e$ is given by  
$H=\case{1}{2m_e}({\bf P}+e{\bf A})^2$.  
However, since the electron is 
confined on the spherical surface, one shows, 
\begin{eqnarray}
H=\frac{1}{2m_eR^2} [{\bf r}\times({\bf P}+e{\bf A})]^2
= \frac{{\omega}_c}{2S} {\bf\Lambda}^2,
\label{hamilu}
\end{eqnarray}
where  ${\bf \Lambda}={\bf r}\times({\bf P}+e{\bf A})$,
${\omega}_c$ is the cyclotron frequency, 
${\bf P}=-i {\bf \nabla}\, $,  
$\nabla\times{\bf A}=B{\bf\hat \Omega}$,
and ${\bf \hat \Omega}={\bf r}/R$.

The components of ${\bf \Lambda}$  obey the commutation relations
$[\Lambda_i,\Lambda_j]=i\epsilon_{ijk}(\Lambda_k-S\Omega_k)$.
The angular momentum operators ${\bf L}={\bf\Lambda}+S{\bf\Omega}$,
and their commutation relations are 
$[L_i,L_j]=i\epsilon_{ijk}L_k$.
Since ${\bf\Lambda}$ is normal to the surface,
we have ${\bf\hat{\Omega}}\cdot {\bf\Lambda}
={\bf\Lambda}\cdot {\bf\hat{\Omega}}=0$,
and ${\bf L}\cdot {\bf\hat{\Omega}}
={\bf\hat{\Omega}}\cdot {\bf L}=S$.
Using these equations, the relation 
$|{\bf\Lambda}|^2=|{\bf L}|^2-S^2$ can be obtained.
Thus the eigenvalues of $|{\bf \Lambda}|^2$ can be deduced 
from the usual angular momentum algebra
$|{\bf\Lambda}|^2=|{\bf L}|^2-S^2=L(L+1)-S^2$, 
$L=S+n, \; n=0,1,2...$,  
and the eigenstates of the Hamiltonian 
are  the eigenstates of $|{\bf L}|^2$
and $L_3 $, and they are 
given by monopole spherical  harmonics. We choose a gauge  field 
${\bf A}=-\case{S}{eR}
\case{(1+\cos\theta)}{\sin\theta}{\bf \hat \varphi}$, 
of which the singularity lies on the north pole 
(we choose a different gauge 
from the one used in Ref.\ \onlinecite{Haldane605}).
The wave functions  at the LLL are given by 
\begin{eqnarray}
u^{S+m} v^{S-m},
\end{eqnarray} 
where  $m=-S, -S+1, \cdots, S$,  and 
\begin{equation}
u =  \cos(\frac{1}{2}\theta)e^{i\varphi},  \, \, \, \, \, 
v =  \sin(\frac{1}{2}\theta).
\end{equation}
All wave functions of the LLs can 
be derived by this way,\cite{wuyang}
and we will not repeat this derivation here.  
In the following, all eigenstates will be obtained by using
projective coordinates, \cite{Fano,LiEsf} and the method developed
in Ref.\ \onlinecite{iengoli}. 

The projective  coordinates are given by
$z=2R\cot\case{\theta}{2} e^{i\varphi}$
and its complex conjugate $\bar z$. 
We will take $R=1/2$ for simplicity.
The measure on the sphere is $\int\case{dxdy}{(1+z\bar z)^2}$.  
The Hamiltonian of Eq.\ (\ref{hamilu})
in  projective coordinates 
is now written by the following formula,\cite{LiEsf}
\begin{equation}
H=\frac{2}{m_e} (1+z\bar z)^2 (P_z+eA_z) (P_{\bar z}+eA_{\bar z}),
\end{equation}
where
\begin{equation}
P_z=-i\frac{\partial}{\partial z}, \;\;\;\
P_{\bar z}=-i\frac{\partial}{\partial \bar z}, \;\;\;\
eA_z=i\frac{\phi}{2}\frac{\bar z}{1+z\bar z}.
\label{hamidue}
\end{equation}
and $\phi$ is the flux 
(in the unit of the fundamental flux $\phi_0$) out of the surface.
Note that the Hamiltonian given by Eq.\ (\ref{hamidue})
(we call this  Hamiltonian as $H^{\prime}$ in the appendix)
is different from the one given by Eq.\ (\ref{hamilu}) by a constant.

The ground states can be determined from the solutions of the equation  
$(P_{\bar z}+eA_{\bar z})\psi=0$, and they are (unnormalized)
\begin{equation}
\psi=\frac{z^l}{(1+z\bar z)^{\frac{\phi}{2}}},
\end{equation}
where $l=0, \cdots , \phi$. At any Landau levels, 
the eigenstates (unnormalized) are given by (from the appendix),
\begin{eqnarray}
\psi_{n,l} & = & [\partial_z+(\frac{B}{2}+1)\partial_z ln g]
[\partial_z+(\frac{B}{2}+2)\partial_z ln g] \cdots \nonumber \\ 
& &  \times [\partial_z+(\frac{B}{2}+n-1)
\partial_z ln g] \psi_{n,l}^{(0)},
\label{allLLs}
\end{eqnarray}
where 
\begin{mathletters}
\begin{eqnarray}
g & = & \frac{1}{(1+z \bar z)^2}, \\
\psi_{n,l}^{(0)} & = &  g^{B/2} \widetilde{\psi}_{n,l}^{(0)}, \\
B & = & \phi/2, \\
\widetilde{\psi}_{n,l}^{(0)} & = & 1,z,\ldots,z^l,\ldots,z^{\phi+2n}.
\end{eqnarray}
\end{mathletters}
Under any finite rotations,  $z$ coordinate is transformed as
$z'=\case{a z+b}{c z+d}$. The rotation matrix 
$R=\left(\begin{array}{cc} a & b \\ c & d \end{array}\right)$ 
is generated by the rotations along the three Cartesian axes,
\begin{mathletters}
\begin{eqnarray}
R_x & = & \frac{1}{\sqrt 2}\left(\begin{array}{cc}
(1+\cos\alpha)^{1/2} & i(1-\cos\alpha)^{1/2} \\
i(1-\cos\alpha)^{1/2} &  (1+\cos\alpha)^{1/2}\end{array}\right), \\
R_y & = & \frac{1}{\sqrt 2}\left(\begin{array}{cc}
(1+\cos\beta)^{1/2} & (1-\cos\beta)^{1/2} \\
-(1-\cos\beta)^{1/2} &  (1+\cos\beta)^{1/2}\end{array}\right), \\
R_z & = & \left(\begin{array}{cc}
\exp(i\gamma/2) & 0 \\
0 & exp(-i\gamma/2)\end{array}\right).
\end{eqnarray}
\end{mathletters}
The rotational invariance of Hamiltonian is shown by the identity:
\begin{equation}
O H(z^{\prime}) O^{-1}=H(z),
\end{equation}
where 
\begin{equation}
O=(\frac{c z+d}{\bar c\bar z+\bar d})^\frac{\phi}{2}.
\end{equation}
The wave function is transformed under rotations as
\begin{equation}
\psi^{\prime} = O\psi(\frac{a z+b}{c z +b}).
\end{equation}
We list some useful relations when we do a finite 
rotation  on a many-body wave function.
\begin{equation}
d(z_i,z_j)=\frac{z_i-z_j}
{\sqrt{1+z_i\bar{z_i}}\sqrt{1+z_j\bar{z_j}}},
\end{equation}
$z_i-z_j$, and $1+z_i\bar z_i$
are transformed under the finite rotation as 
\begin{mathletters}
\begin{eqnarray}
d(z_i^{\prime},z_j^{\prime}) & = & 
(\frac{c z_i+d}{\bar c\bar z_i+\bar d})^\frac{1}{2}
(\frac{c z_j+d}{\bar c\bar z_j+\bar d})^\frac{1}{2}
d(z_i,z_j), \\
z_i'-z_j' & = & \frac{z_i-z_j}{(cz_i+d)(cz_j+d)}, \\
1+\bar z_i'\bar z_i' & = & 
\frac{1+z_i\bar z_i}{(cz_i+d)(\bar c\bar z+\bar d)}.
\end{eqnarray}
\end{mathletters}
Finally, the angular momentum operators for $N$ electrons are
\begin{mathletters}
\label{multiang}
\begin{eqnarray}
J_x & = & \sum_{i=1}^N J_x(i) \nonumber \\
& = & \frac{1}{2}
\sum_{i=1}^N[(1-z_i^2)\frac{\partial}{\partial z_i}
-(1-\bar z_i^2)\frac{\partial}{\partial\bar z}
+\frac{\phi}{2}(z_i+\bar z_i)],
\\
J_y & = & \sum_{i=1}^N J_y(i)  \nonumber \\
& = &  \frac{i}{2}\sum_{i=1}^N[(1+z_i^2)
\frac{\partial}{\partial z_i}
+(1+\bar z_i^2)\frac{\partial}{\partial\bar z}
+\frac{\phi}{2}(\bar z_i-z_i)],
\\
J_z & = & \sum_{i=1}^N J_z(i) \nonumber \\
& = & \sum_{i=1}^N(z_i\frac{\partial}{\partial z}-
\bar z_i\frac{\partial}{\partial\bar z}-\frac{\phi}{2}).
\end{eqnarray}
\end{mathletters}

\section{Projections and Angular Momentums in  the LLL}

The FQH state is restricted to the LLL. 
In this section, we will discuss 
briefly how to project states to the LLL
on the sphere (see Ref.\ \onlinecite{Fano}, and for 
the detailed discussions in the case of 
a plane or a disk, see Ref.\ \onlinecite{GirvinJach}), 
and how to find the eigenstates of angular momentums when the
particles are restricted to the LLL.  
Note that the construction of  
the CF wave functions involves the
higher LLs,  we need to project the wave functions  to the LLL
(see Sec.\ \ref{sec:cf}).

The normalized states with flux $\phi$ in the LLL are  
\begin{equation}
|l>= [\frac{(\phi+1)!}{2\pi l!(\phi-l)!}]^{1/2}
\frac{z^l}{(1+z\bar z)^{\phi/2}}, 
\end{equation}
and  $l=0, 1, 2, \cdots , \phi$.
The projection operator to the LLL 
is $P=\sum_l|l><l|$, and it can be
written also in the following form, 
\begin{mathletters}
\begin{eqnarray}
P\psi (z, \bar z) & = &  
\int \frac{dw d \bar w}{(1+w \bar w)^2} G(z, w)
\psi (\omega, \bar \omega),
\\
G(z, w) & = & \frac{\phi +1}{2\pi} \frac{(1+z\bar w)^{\phi}}
{(1+z \bar z)^{\phi /2} (1+w \bar w)^{\phi /2}}.
\end{eqnarray}
\end{mathletters}
For the many-body wave functions,  $P$ (or $G$)
is equal to $\prod_{i=1}^N P_i$ (or $\prod_{i=1}^N  G_i$)
where $P_i$ is the projection operator of the $\text{i-th}$ particle
and $N$ is the number of particles.

If the state is not in the LLL, the anti-holomorphic 
coordinate $\bar z$  will appear. Typically, it appears as
\begin{equation}
\psi= \frac{\bar z^i z^{i+l} }{(1+z\bar z)^{(\phi/2)+j }},
\end{equation}
and $P\psi$  is equal to
\begin{equation}
\frac{(\phi +1)!(l+i)!(\phi +j-l-i)!}{l!(\phi -l)! (\phi +j+1)!}
\frac{z^l}{(1+z\bar z)^{\phi/2}}.
\end{equation}
On the sphere, if the interactions between electrons are rotationally
invariant, the eigenstates of the many-body Hamiltonian
should be also the eigenstates of rotational  operators $J^2$ and $J_z$. 
The FQH ground states  on the sphere are rotationally
invariant and are non-degenerated.
In order to find the ground states, we can
thus use the rotational invariant states to diagonalize
the Hamiltonian. As the number of 
all possible rotational invariant states is much less than
the number of all possible states, it is thus much easier
to find the ground states by using the rotational invariant states 
to diagonalize the Hamiltonian than by using all possible states.
It could be also interesting to find the eigenstates of
$J^2 \not= 0$ (which are not rotationally invariant). 
The excited states in the FQH are not rotationally invariant.
For Fermi-liquid-like  systems in a half-filled Landau level, one 
can have ground states which are 
not rotationally invariant.\cite{rezayi}

Now we are going to find the many-body wave functions
on the LLL which are the eigenstates of $J^2$ and $J_z$.
In the LLL, the many-body wave functions $\Psi$ have the form
\begin{equation}
\Psi=\prod_{i=1}^N \frac{1}{(1+z_i\bar z_i)^{\frac{\phi}{2}}}
F(z_1,z_2, \cdots, z_N),
\end{equation}
where $F(z_1,z_2, \cdots, z_N)$ is an  anti-symmetric
holomorphic function.
When $J_{+}=J_x +iJ_y, J_{-}=J_x -iJ_y, J_z$ act on $\Psi$, 
we have
\begin{mathletters}
\begin{eqnarray}
J_{-}\Psi & = &
\prod_{i=1}^N \frac{1}{(1+z_i\bar z_i)^{\frac{\phi}{2}}}
\sum_{i=1}^N \frac{\partial}{\partial z_i} F,   \\
J_{+}\Psi & = & \prod_{i=1}^N 
\frac{1}{(1+z_i\bar z_i)^{\frac{\phi}{2}}}
\sum_{i=1}^N (-z_i^2\frac{ \partial}{\partial z_i}+\phi z_i)F, \\
J_z \Psi & = & \prod_{i=1}^N  
\frac{1}{(1+z_i\bar z_i)^{\frac{\phi}{2}}}
[(\sum_{i=1}^N  z_i\frac{\partial}{\partial z_i})
-\frac{N\phi}{2}]F.
\end{eqnarray}
\end{mathletters}
Thus the projected ${\bf J}$ operators are:
\begin{mathletters} 
\label{LLsang}
\begin{eqnarray}
J_{-}^{\prime} & = &\sum_{i=1}^N \frac{\partial}{\partial z_i},   \\
J_{+}^{\prime} & = & \sum_{i=1}^N -z_i^2
\frac{ \partial}{\partial z_i}+\phi z_i,  \\
J_z^{\prime} & = & (\sum_{i=1}^N
z_i \frac{\partial}{\partial z_i})-\frac{N\phi}{2},
\end{eqnarray}
\end{mathletters}
where they  act only on $F$. 
The angular momentum eigenstates of 
the many-body wave functions restricted 
to the LLL can be obtained by solving 
\begin{mathletters} 
\label{angeqsa}
\begin{eqnarray}
J_{-}^{\prime}F(-J)& = & 0, \\ 
J_z^{\prime}F(-J) & = &  -JF(-J),
\end{eqnarray}
\end{mathletters}
where $F(-J)$ is the lowest weight eigenstate with weight $-J$.   
Other states can be obtained by 
applying $J_{+}^{\prime}$ repeatedly on $F(-J)^{\prime}$.
Eq.\ (\ref{angeqsa}) leads to
\begin{mathletters}
\label{nang}
\begin{eqnarray}
\sum_{i=1}^N z_i\frac{ \partial}{\partial z_i} F(-J)
& = & (\frac{N\phi}{2}-J)F(-J),  \\
\sum_{i=1}^N \frac{\partial}{\partial z_i} F(-J)
& = & 0 .
\end{eqnarray}
\end{mathletters}
The first equation in Eq.\ (\ref{nang})
means that $F(-J)$ is a homogeneous polynomial
with degree $\case{N\phi}{2} -J$. 
As $F(-J)$ is an anti-symmetric function of holomorphic coordinates, 
it can be factorized 
as $F(-J)=\prod_{i<j}^N (z_i-z_j) F^{\prime}(-J)$.
One can check that 
\begin{mathletters}
\label{nang1}
\begin{eqnarray}
J_{-}^{\prime}F(-J) &
= & \prod_{i<j}^N (z_i-z_j) J_{-}^{\prime}F^{\prime}(-J),  \\
J_{z}^{\prime}F(-J)
& = & \frac{N(N-1)}{2}F(-J)
+\prod_{i<j}^N (z_i-z_j)  J_{z}^{\prime} F^{\prime}(-J).
\end{eqnarray}
\end{mathletters}
Thus $F^{\prime}(-J)$ is a symmetric function with degree 
$L= \case{N\phi}{2}-J-\case{N(N-1)}{2}$,  and the power of every
coordinate in $F^{\prime}(-J)$ shall be less or equal than 
$\phi^{\prime}$ where $\phi^{\prime}=\phi -(N-1)$. 
By using Eq.\ (\ref{nang})  and Eq.\ (\ref{nang1}),  one finds that
$F^{\prime}(-J)$ satisfies the conditions:
\begin{mathletters}
\label{eqnum1}
\begin{eqnarray}
J_{-}^{\prime}F^{\prime}(-J) & = & 0 ,  \\
J_{z}^{\prime}F^{\prime}(-J)
& = & (\frac{N\phi^{\prime}}{2}-J)F^{\prime}(-J).
\end{eqnarray}
\end{mathletters}
Define symmetric polynomials $\sigma_i$:
\begin{equation}
P(z_i)=\prod_{i=1}^N (z-z_i)=\sum_{i=0}^N (-1)^i \sigma_i z^{N-i},
\end{equation}
where 
\begin{equation}
\sigma_0=1, \, \,  \sigma_1=\sum_{i=1}^N  z_i, 
\, \, \cdots, \, \,  \sigma_N=\prod_{i=1}^N  z_i.
\end{equation}
$F^{\prime}$ can be expanded as 
\begin{equation}
\sum_{s_i}  C(s_i)\prod_{i=1}^N \sigma^{s_i}_i,
\end{equation}
where $s_i$  is a non-negative integer. 
By using Eq.\ (\ref{eqnum1}), we get 
equations  which  $C(s_i)$ and $s_i$ must obey.  One of them is
\begin{eqnarray}
\sum_{i=1}^N  i s_i  =L=  \frac{N\phi^{\prime}}{2}-J.  
\label{eqnum2}
\end{eqnarray}
The condition
\begin{equation}
\sum_{i=1}^N  s_i  \leq  \phi^{\prime} 
\end{equation}
must be satisfied in order that the wave function is normalizable.
$C(s_i)$ shall also satisfy the equation 
\begin{equation}
J_{-}^{\prime}F^{\prime}(-J) = \sum_{s^{\prime}_i} 
C^{\prime}(s^{\prime}_i)
\prod_{i=1}^N \sigma^{s^{\prime}_i}_i=0, 
\end{equation}
where $C^{\prime}(s^{\prime}_i)$ is a linear combinations of 
$C(s_i)$,  and it shall be equal to $0$.
Because $[J_z \, , J_{-}]=-J_{-}$,  $s^{\prime}_i$ shall satisfy 
\begin{equation}
\sum_{i=1}^N  is^{\prime}_i  =L-1. 
\label{eqnum3}
\end{equation} 
Apparently $\sum_{i=1}^N  s^{\prime}_i$ shall be 
also less or equal than $\phi^{\prime}$.  
The number of linear independent  solutions for $C(s_i)$ is 
equal to the number of solutions of Eq.\ (\ref{eqnum2})
minus the number of solutions of Eq.\ (\ref{eqnum3}), and it
is also equal to  $M(J,N,\phi^{\prime})$, which is the number
of states with spin $J$.

The generation function for the number of solutions of 
Eq.\ (\ref{eqnum2}) or Eq.\ (\ref{eqnum3}) is 
\begin{eqnarray}
G(t,q)=\prod_{i=1}^N \frac{1}{1-tq^i}.
\end{eqnarray}
The number of solutions of Eq.\ (\ref{eqnum2}) 
is equal  to the sum of the coefficient of term 
$t^jq^L$ with $0 \leq j \leq \phi^{\prime}$ in $G(t,q)$. 
Thus $M(J,N,\phi^{\prime})$ is equal to
\begin{equation}
\oint \oint \frac{dt}{2\pi i t} \frac{dq}{2\pi i q}
G(t,q) (\frac{1}{q^L}-\frac{1}{q^{L-1}})
\sum_{i=0}^{\phi^{\prime}}\frac{1}{t^i}.
\end{equation}
We can also use a generation function of one variable,\cite{Andrews}
\begin{equation}
G(t)=\frac{\prod_{k=1}^{N+\phi'} (1-t^k)}{\prod_{k=1}^N 
(1-t^k)\prod_{k=1}^{\phi'}(1-t^k)}.
\end{equation}
The number of solutions of Eq.\ (\ref{eqnum2}) is then 
given by the coefficient of $t^L$ of function $G(t)$.  Thus 
\begin{equation}
M(J,N,\phi^{\prime})=\oint \frac{dt}{2\pi i t} G(t)\frac{1-t}{t^L}.
\end{equation}
The asymptotic behavior of 
$M(J,N,\phi^{\prime})$ can be obtained by using
the steepest descent method.\cite{itzykson1} 
When $L$ and $\phi^{\prime}$ are both large, 
$M(J,N,\phi^{\prime})$ is equal to
$exp(S(J,N, \phi^{\prime}))$ approximately, and
$exp(S(J,N, \phi^{\prime}))$ is determined by the following equations,
\begin{mathletters}
\begin{eqnarray}
L & = & -\frac{exp(-\rho)}{1- exp(-\rho )}+
\frac{1}{\rho^2}
(-\int_0^{(N+\phi^{\prime})\rho}
+\int_0^{N\rho}  \nonumber \\
& & +\int_0^{\phi^{\prime}\rho}) 
du \frac{uexp (-\rho u)}{1-exp (-\rho u)},   \\ 
S(J,N, \phi^{\prime}) & = & 
L \rho + \ln (1-exp(-\rho )) +
\frac{1}{\rho}(\int_0^{(N+\phi^{\prime})\rho}-
\int_0^{N\rho}   \nonumber \\ & & 
+\int_0^{\phi^{\prime}\rho}) du \ln [1-exp (- u)]. 
\end{eqnarray}
\end{mathletters}
We list the number of rotational invariance 
states at various fillings in Table \ \ref{table1}.

\section{Hierarchical wave functions}

In this section, we will discuss the construction 
of  the hierarchical wave 
functions.\cite{PrangeGirvin,Haldane605,Halperin1}
The quasiparticles satisfy fractional statistics, and 
the condensation of quasiparticles 
gives rise to the FQH state with  $\nu \not= 1/m$. 

Define
\begin{equation}
\Psi_m=\prod_{i<j}^N(u_iv_j-u_jv_i)^m, 
\end{equation}
where $m$ is a positive integer. 
For $\nu=\case{1}{m}$ with $m$ being an old positive integer, 
the FQH wave function  will be $\Psi_m$   (the Laughlin wave function).
The flux $\phi$ is equal to $\phi_m=m(N-1)$. \cite{Haldane605}     
Or in the projective coordinates, it is
\begin{equation}
\Psi_m=\prod_{i<j}^Nd(z_i,z_j)^m .
\end{equation}
The Laughlin wave function with the presence 
of   quasiparticle excitations is given by acting
the quasiparticle excitation operators on the original 
Laughlin wave function. 
The quasiparticle excitation operator is given by
\begin{mathletters}
\begin{eqnarray}
A^{\dagger}(\alpha,\beta) & = & 
\prod_{i=1}^N(\beta u_i-\alpha v_i), \;\;\;\;\;\;\; 
\text{(quasihole)} \\
A(\alpha,\beta)& = & \prod_{i=1}^N
(\bar\beta\frac{\partial}{\partial u_i}-
\bar\alpha\frac{\partial}{\partial v_i}),  \;\;\;\;\;\;\;
\text{(quasielectron)}
\end{eqnarray}
\end{mathletters}
where $\alpha=\cos \case{\theta}{2} e^{i\varphi}$, 
and  $\beta=\sin \case{\theta}{2}$ are the quasiparticle coordinates. 
In the projective coordinates, the operators of the quasihole excitation 
and the quasielectron excitation are  given in the following form,
\begin{mathletters}
\label{uno}
\begin{eqnarray}
A^{\dagger}(\omega, \bar \omega ) \Psi_m(z_i)
& = & \prod_{i=1}^N d(z_i, \omega)\Psi_m(z_i), \\
A(\omega, \bar \omega ) \Psi_{m}(z_i) & = & 
\frac{1}{(1+\omega \bar\omega)^{N/2}}
\prod_{i=1}^N \frac{1}{(1+z_i\bar z_i)^{\frac{\phi_m-1}{2}}}
\nonumber \\
& & \times \prod_{i=1}^N [(1+z_i\bar\omega ) 
\partial_{z_i}-\phi_m \bar\omega ]F_{m},
\end{eqnarray}
\end{mathletters}
where $\omega \,\, ,  \bar \omega$ is the 
projective  coordinates of the quasiparticle,
and $F_m(z_i)=\prod_{i<j}^N (z_i-z_j)^m $.  
The flux $\phi$ in the presence of a quasielectron
(quasihole) is $\phi_m -1$ ($\phi_m +1$).

The slightly entangled appearance of $A(\omega)$ 
hides, indeed, a form which is analogous to ${A}^{\dagger}(\omega)$.  
To unveil the similarities,  one can show that
\begin{equation}
P(\phi_m-1,z_i) \prod_{i=1}^N d(\bar z_i, \bar \omega) \Psi_m
\end{equation}
gives the wave function of the Laughlin state in the presence
of a quasihole as that in Eq.\ (\ref{uno}).  
$P(\phi,z_i)$  (here $\phi=\phi_m-1$)  
projects the wave function 
to  the LLL with flux $\phi $ with respect
to coordinates $z_i$.  Thus the construction of
the hierarchical wave functions 
due to the condensation of quasielectrons
will {\it naturally} involve higher Landau levels as 
in the case of the CF wave functions (see the next section).

Instead using $A(\omega, \bar \omega )\Psi_m (z_i)$,
we can also create a quasielectron excitation using 
$\Psi_{m-2}A D$,\cite{greiter,hard} 
where $D=[\Psi_{1}(z_i)]^{2}$.  $A D$ is here equal to  
$P(\phi_2-1)\prod_{i=1}^N  d(\bar z_i, \bar \omega) D$.
We call $(\Psi_{1})^{m-2}A D$  as a wave function 
by the hard core construction.  

In the case of many quasiparticle excitations, 
the operators of excitations are
\begin{mathletters}
\begin{eqnarray}
A^{\dagger}_{N_q} & = & 
\prod_{j=1}^{N_q} A^{\dagger}(\omega_j, \bar \omega_j ),  \\
A_{N_q} & =  & 
\prod_{j=1}^{N_q} A(\omega_j, \bar \omega_j ).
\end{eqnarray}
\end{mathletters}
When $A_{N_q}$ acts on $\Psi_m$,  one can show
\begin{mathletters}
\begin{eqnarray}
A_{N_q} \Psi_m & = & \prod_{j=1}^{N_q} 
\frac{1}{(1+\omega_j \bar\omega_j )^{N/2}}
\prod_{i=1}^N  \frac{1}{(1+z_i\bar z_i)^{\frac{\phi_m-N_q}{2}}}
\nonumber \\
& & \times A^{\prime}_{N_q}F_m(z_i),    
\end{eqnarray}
\end{mathletters}
where
\begin{mathletters}
\label{prodoper}
\begin{eqnarray}
A^{\prime}_{N_q} & =& \prod_{j=1}^{N_q} A^{\prime}(j)
\nonumber \\ 
& = &
\prod_{i=1}^N [(1+z_i\bar\omega_{N_q} ) 
\partial_{z_i}-(\phi_m -N_q+1) \bar\omega_{N_q}]
\nonumber \\
& & \times 
\prod_{i=1}^N [(1+z_i\bar\omega_{N_q-1} ) 
\partial_{z_i} \nonumber \\
& & -(\phi_m -N_q+2) \bar\omega_{N_q-1}]
\nonumber \\
& &  \cdots  \cdots \cdots \cdots
\nonumber \\
& & \times
\prod_{i=1}^N [(1+z_i\bar\omega_{j} ) 
\partial_{z_i}-(\phi_m -j+1) \bar\omega_{j}]
\nonumber \\
& &  \cdots  \cdots \cdots \cdots
\nonumber \\
& & \times
\prod_{i=1}^N [(1+z_j\bar\omega_1 ) 
\partial_{z_j}-\phi_m \bar\omega_1 ].
\end{eqnarray}
\end{mathletters}
One shall be careful about the ordering of 
$A^{\prime}(j)$ in Eq.\ (\ref{prodoper}).  
$A^{\prime}(j)$ in Eq.\ (\ref{prodoper}) is defined as
\begin{equation}
A^{\prime}(j)=\prod_{i=1}^N [(1+z_i\bar\omega_{j}) 
\partial_{z_i}-(\phi_m -j+1) \bar\omega_{j}].
\end{equation}
As in the case of a single quasihole excitation,
the wave function in the presence of $N_q$
quasiholes can be also written as
\begin{equation}
P(\phi )\prod_{i=1}^N \prod_{\alpha=1}^{N_q}
d(\bar z_i, \bar \omega_{\alpha})\Psi_{m},
\end{equation}
where $\phi =\Phi_m -N_q$.

To construct the hierarchical wave functions, we shall 
normalize the Laughlin wave functions in the presence of
quasiparticles. One can show that 
$\Psi_{\text{e,q}}=[\Psi_1(\omega_\alpha )]^{1/m} 
 A^{\dagger}_{N_q}\Psi_{m}$,
$\Psi_{\text{e,q}}= 
[\Psi_1(\omega_\alpha )]^{1/m}A_{N_q}\Psi_{m}$,  
or for the hard core constructed wave function,
$[\Psi_1(\omega_\alpha )]^{1/m} \Psi_{m-2}A D$,
are normalized.\cite{arovas}
The hierarchical states are obtained if the quasiparticles
are also condensed to the Laughlin states.
The wave function of quasiholes is 
$\Psi_q=[\Psi_1(\bar \omega_{\alpha}]^{p_2+\case{1}{m}} $,
the wave function of quasielectrons is 
$\Psi_q=[\Psi_1(\omega_{\alpha})]^{p_2- \case{1}{m}}$, 
and $p$ is a positive even integer.
Quasiparticles satisfy fractional statistics,\cite{frast}
and the wave functions here are in singular 
gauge which shows  fractional statistics explicitly. 
The hierarchical wave function 
for the electrons is then given by the following formula,
\begin{equation}
\int \prod_{\alpha=1}^{N_q}
\frac{d\omega_{\alpha}d\bar \omega_{\alpha}}{
(1+\omega_{\alpha}\bar \omega_{\alpha})^2}
\Psi{\text{e,q}} \Psi_q,
\end{equation}
or explicitly, 
\begin{eqnarray}
\label{trew}
\Psi_e (m, p)&=& \int \prod_{\alpha =1}^{N_q}
\frac{d\omega_{\alpha}d\bar \omega_{\alpha}}{
(1+\omega_{\alpha}\bar \omega_{\alpha})^2}
\Psi_m (z_i) \prod_{i=1}^N \prod_{\alpha =1}^{N_q} 
d(z_i, \omega_{\alpha}) \nonumber \\
& & \times
|\Psi_1(\omega_{\alpha})|^{2/m} \Psi_{p_2}(\bar \omega_{\alpha})
\end{eqnarray}
is the hierarchical wave function due to 
the condensation of quasiholes, and
the hierarchical wave functions due to 
the condensation of quasielectrons by the non-hard core
construction and the hard core construction
are given in the following formulas, 
\begin{mathletters}
\begin{eqnarray}
\Psi_e (m, -p)&=& P(\phi, z_i) \int \prod_{\alpha =1}^{N_q}
\frac{d\omega_{\alpha}d\bar \omega_{\alpha}}{
(1+\omega_{\alpha}\bar \omega_{\alpha})^2}\Psi_m (z_i)  
\nonumber  \\
& & \times \prod_{i=1}^N
\prod_{\alpha=1}^{N_q} d(\bar z_i, \bar \omega_{\alpha}) 
\Psi_{p_2}(\omega_{\alpha}),   \\
\Psi_{\text{e,hard}}(m, -p) 
&=& \Psi_{m-2}(z_i) P(\phi_2 -N_q,z_i)
\int \prod_{\alpha =1}^{N_q}
\frac{d\omega_{\alpha}d\bar \omega_{\alpha}}{
(1+\omega_{\alpha}\bar \omega_{\alpha})^2} \nonumber \\
& & \times \Psi_2(z_i) \prod_{i=1}^N \prod_{\alpha =1}^{N_q}
d(\bar z_i, \bar \omega_{\alpha})
\Psi_{p_2}(\omega_{\alpha}).
\end{eqnarray}
\end{mathletters}
We also require the wave functions above being rotationally
invariant. This requirement leads to
\begin{mathletters}
\label{quaselw}
\begin{eqnarray}
m(N-1)+\xi_2 N_q& = & \phi,  \\
p_2(N_q-1)& = & N.
\end{eqnarray}
\end{mathletters}
$\xi_2=\pm 1$ in the case of the condensation of quasiholes
and quasielectrons respectively.  And the Landau level filling 
fraction $\nu$ is equal to 
\begin{equation}
\frac{1}{m+\frac{1}{\xi_2p_2}}.
\end{equation}
For $m=1$ and $\xi_2=1$, the filling 
$\nu =\case{1}{m+\case{1}{p_2}}$ is
equal to the filling of the charge conjugate state, 
$1-\case{1}{p_2+1}$.  Actually,  the 
wave function $\Psi_e (m, p)$ is also  the charge conjugate of the
Laughlin wave function at filling 
$\nu =\case{1}{p_2+1}$,  and this shows that the construction 
of the wave function is consistent with physical picture.
When $m\not= 1$,  we notice that, in the formula for $\Psi_e (m, p)$, 
we can not do the integration exactly due to the term 
$|\Psi_1(\omega_{\alpha})|^{2/m}$.
We can approximate the trial wave function $\Psi_e (m, p)$ 
by omitting $|\Psi_1(\omega_{\alpha})|^{2/m}$,   
and it becomes 
\begin{eqnarray}
\Psi_e (m, p) &  \approx & \int \prod_{\alpha =1}^{N_q}
\frac{d\omega_{\alpha}d\bar \omega_{\alpha}}{
(1+\omega_{\alpha}\bar \omega_{\alpha})^2}
\Psi_m (z_i)   \nonumber \\
& & \times \prod_{i=1}^N \prod_{\alpha =1}^{N_q} 
d(z_i, \omega_{\alpha}) \Psi_{p_2}(\bar \omega_{\alpha}).
\label{qusaieapp}
\end{eqnarray}
The wave function written in Eq.\ (\ref{qusaieapp}) 
is still rotationally invariant,\cite{greiter,LiEsf},
and  we are  able to integrate it.
When $m=1$, the formula for $\Psi_e (m, p)$ in
Eq.\ (\ref{trew}) is integrable. 
When $m=1$, we find that 
the overlap between the wave functions given 
in Eq.\ (\ref{trew}) and Eq.\ (\ref{qusaieapp})
is excellent for a small number of electrons.
In Ref.\ \onlinecite{greiter}, it was also found that 
the overlapping of the  wave functions 
given by Eq.\ (\ref{qusaieapp})
with the exact ground state of the FQH is
all most equal to one for a small number of electrons.
We note that  the wave functions calculated in Sec.\ \ref{fileoverl}
are  based on  the formulas written in 
Eq.\ (\ref{quaselw}) and Eq.\ (\ref{qusaieapp}).

In the formula for $\Psi_e (m, -p)$ or
$\Psi_{\text{e,hard}}(m, -p)$, 
we note that one can do the integration first,
and then  the projection, or vice versa.
In Sec.\ \ref{fileoverl}, the overlap 
between these two wave functions will be calculated
for a small number of electrons 
and it is found that the overlap is all most  equal to one.

We will call the above hierarchical states as the $2nd$-level
hierarchical states, and the Laughlin states as the $1st$-level
hierarchical states. The higher-level hierarchical states can 
be built in the similar way.\cite{Read,wenblok,LiEsf,greiter} 
We denote the $k'th$-level hierarchical states by
$(p_1, \xi_2p_2, \xi_3p_3, \cdots , \xi_kp_k)$,
where $p_1$ is an old positive integer, $p_i \, , i\not= 1$
are even positive integers, and  $\xi_i =\pm $ indicate
the quasihole condensation  and quasielectron condensation
from parent states.

For the higher-level hierarchical wave functions involving
the condensation of quasielectrons, we can make 
a {\it further} simplification. 
We take $(p_1, -p_2, \\ -p_3)$ as an  example. The wave function 
for this state is 
\begin{eqnarray}
\Psi_e &=& P(\phi, z_i) \int \prod_{\alpha =1}^{N_2}
\frac{d\omega^1_{\alpha}d\bar \omega^1_{\alpha}}{
(1+\omega^1_{\alpha}\bar \omega^1_{\alpha})^2}\Psi_m (z_i)  
 \prod_{i=1}^{N_1} \prod_{\alpha=1}^{N_2} 
d(\bar z_i, \bar \omega^1_{\alpha})\nonumber \\
& & \times P(N_1, \omega^1_{\alpha})\int \prod_{\alpha =1}^{N_3}
\frac{d\omega^2_{\alpha}d\bar \omega^2_{\alpha}}{
(1+\omega^2_{\alpha}\bar \omega^2_{\alpha})^2}
\Psi_{p_2}(\omega^1_{\alpha}) \nonumber \\
& &  \times  \prod_{\alpha=1}^{N_2} \prod_{\beta=1}^{N_3} 
d(\bar \omega^1_{\alpha},
\bar \omega^2_{\beta})\Psi_{p_3}(\omega^2_{\alpha}),
\end{eqnarray}
where $N_1$ is the number of electrons, $N_2$ is the number
of quasielectrons of the Laughlin state  $(p_1)$, 
$N_3$ is the number of quasielectrons of the hierarchical state
$(p_1, -p_2)$,  $\omega^1_{\alpha}$ and  $\omega^2_{\alpha}$ 
are the coordinates of quasiparticles of the two types  
respectively. We can  prove that 
$P(N_1, \omega^1_{\alpha})$ can be {\it drooped}
inside the formula. Thus the wave function
can be written as 
\begin{eqnarray}
\Psi_e &=& P(\phi, z_i) \int \prod_{\alpha=1}^{N_2}
\prod_{\beta=1}^{N_3}
\frac{d\omega^1_{\alpha}d\bar \omega^1_{\alpha}}{
(1+\omega^1_{\alpha}\bar \omega^1_{\alpha})^2}
\frac{d\omega^2_{\beta}d\bar \omega^2_{\beta}}{
(1+\omega^2_{\beta}\bar \omega^2_{\beta})^2}
\nonumber \\
& & \times \Psi_m (z_i) \Psi_{p_2}(\omega^1_{\alpha})
\Psi_{p_3}(\omega^2_{\alpha})  \nonumber \\
& & \times
\prod_{i=1}^{N_1}\prod_{\alpha=1}^{N_2} 
d(\bar z_i, \bar \omega^1_{\alpha})
\prod_{\alpha=1}^{N_2}\prod_{\beta=1}^{N_3} 
d(\bar \omega^1_{\alpha},\bar \omega^2_{\beta}).
\label{readlike}
\end{eqnarray}
The  wave function in Eq.\ (\ref{readlike}) is quite similar to
the wave function constructed in Ref.\ \onlinecite{Read},
\begin{eqnarray}
\Psi_e &=& P(\phi, z_i) \int \prod_{\alpha=1}^{N_2}
\prod_{\beta=1}^{N_3}
\frac{d\omega^1_{\alpha}d\bar \omega^1_{\alpha}}{
(1+\omega^1_{\alpha}\bar \omega^1_{\alpha})^2}
\frac{d\omega^2_{\beta}d\bar \omega^2_{\beta}}{
(1+\omega^2_{\beta}\bar \omega^2_{\beta})^2}
\nonumber \\
& & \times \Psi_m (z_i) \Psi_{p_2}(\omega^1_{\alpha})
\Psi_{p_3}(\omega^2_{\alpha})  \nonumber \\
& & \times
\prod_{i=1}^{N_1} \prod_{\alpha=1}^{N_2}  
\frac{1}{d( z_i,  \omega^1_{\alpha})}
\prod_{\alpha=1}^{N_2} \prod_{\beta=1}^{N_3} 
\frac{1}{d(\omega^1_{\alpha}, \omega^2_{\beta})}.
\label{readhier}
\end{eqnarray}
However it  is difficult to handle Eq.\ (\ref{readhier})  in the
practical calculation due to  the singularities.

Finally by requiring the rotational invariance of the wave function
(\ref{readlike}) or (\ref{readhier}), one gets
\begin{eqnarray}
p_1(N_1-1)- N_2& = & \phi, \nonumber \\
N_1-p_2(N_2-1)+N_3& = & 0, \nonumber \\
N_2-p_3(N_3-1) & = & 0,
\label{level3}
\end{eqnarray}
and Eq.\ (\ref{level3}) implies that the filling of 
the FQH state is equal to
\begin{equation}
\frac{1}{p_1+\frac{1}{p_2+\frac{1}{p_3}}}.
\end{equation}
We point out that the wave function  
proposed in  Ref.\ \onlinecite{Read} had been
also constructed on the torus.\cite{torus} 
It would be very interesting  if we can generalize
the construction of the wave function (\ref{readlike}) 
to the torus.

\section{Composite Fermion wave functions}
\label{sec:cf}

The CF theory of the FQHE has  
significantly advanced the understanding of the 
FQHE recently.\cite{Jain3} The FQHE is due to
the integer QHE of the CFs, where a 
CF  is the bound state of  
an electron and an even number of vortices.
We will discuss  in this section how to 
calculate the CF wave functions in our framework.

Jain proposed that all trial wave functions of the FQHE
(note again in this paper the spin is polarized)
can be obtained by using two operations, $D$ and $C$, respectively
composite fermionization and charge conjugation,
on the wave functions  of the integer 
QHE of the CFs. For example,
the trial wave function of electrons at $\nu =n/(2n+1)$ can be written 
as $PD\chi_n$, where $\chi_n$ is the wave function 
of the CFs  which fill completely the first $n$ Landau levels
with flux $\phi^*$ ($P$ is the projection operator 
to the LLL as in the previous sections).
The flux of the state $PD\chi_n$ is equal to
$2(N-1)+\phi^*$ where $\phi^* =\frac{N}{n} -n$.
We can also use $\Psi_1 P(\phi -N+1) \Psi_1 \chi_n$
as the trial wave function and we call this wave function 
as the  wave function by the hard core construction. 
The charge conjugation of 
$PD\chi_n$ (or $\Psi_1 P(\phi -N+1) \Psi_1 \chi_n$)
is then trial wave function 
at $\nu =1 -\case{n}{2n+1}=\case{n+1}{2n+1}$. 
The trial wave function  at other fillings 
can be obtained by acting  repeatedly  $D$ and $C$ 
on $PD\chi_n$ ($\Psi_1 P(\phi -N+1) \Psi_1 \chi_n$)
(each state can be obtained only in a unique way
in this picture except the ordering of operator $P$).

$\chi_n$ is given by the determinant
$\chi_n= \det (\psi_{s,k}(z_i))$, 
where $s=0,1, \cdots, n-1$,
$k=\phi^* +2s+1$, $i=1, 2, \cdots, N$, 
$N=n\phi^* +n^2$. $\det (\psi_{s,k}(z_i))$ can be simplified 
and it is given by the following formula,
\begin{eqnarray} 
\label{determinante}\chi_{n} & = &  
\left| \begin{array}{cccccc}
              1     &      1        &  .  & .  &  .  &  1 \\
            z_{1} &      z_{2}    & .  & .  &  . &  z_{N}\\
              z_{1} &      z_{2}    & .  & .  &  . &  z_{N}\\
            \vdots &   \vdots    
& \vdots  &  \vdots  & \vdots  & \vdots \\
            z_{1}^{N^{\prime}-1} &   z_{2}^{N^{\prime}-1}
  &  .  & .  &  . & z_{N}^{N^{\prime}-1}\\
   \bar z_1  &  \bar z_2  & . & . & . & \bar z_N \\
    \bar z _{1}z_{1}  & \bar z _{2}z_{2}  & .  & .  &  . &  
\bar z_{N}z_{N}\\
             \vdots &   \vdots    & \vdots  
&  \vdots  & \vdots  & \vdots \\
\bar z_{1}z_{1}^{N^{\prime}-1}  & \bar z _{2}z_{2}^{N^{\prime}-1}
  & .  & .  &  . & \bar z _N z_N^{N^{\prime}-1} \\
            \vdots &   \vdots    & \vdots  
&  \vdots  & \vdots & \vdots\\
\bar z_1^{n-1} & \bar z_2^{n-1} &  . & . & . & \bar z_N^{n-1} \\
\bar z_{1}^{n-1}z_{1} & \bar z_{2}^{n-1}z_{2} & . & . & . & 
\bar z_N^{n-1}z_N \\
          \vdots &   \vdots    & \vdots  
&  \vdots  & \vdots  & \vdots\\
\bar z_1^{n-1}z_{1}^{N^{\prime}-1} 
& \bar z_{2}^{n-1}z_{2}^{N^{\prime}-1}
 & . & . & . & \bar 
z_N^{n-1}z_N^{N^{\prime}-1}
\end{array} \right |                      \nonumber \\
& & \times \prod _{j=1}^{N}
\frac{1}{(1+z_j\bar z_j)^{\frac{\phi^*}{2}+n-1}}, 
\end{eqnarray}
where $ N^{\prime}= N/n =\phi^* +n$. 
We divide $N$ electrons into $n$ groups. 
The set of the original coordinates $z_i$ 
can be mapped to $z_{s,k}$ with $s=0, 1, \cdots, n-1$,
$k=1, 2, \cdots, N^{\prime}$.  The determinant  is proportional to
\begin{eqnarray}
\chi_n &= & {\bf \text{AN}} 
\prod_{s=0}^{n-1} [ e^{N^{\prime}}_s ]^s \Psi_{s,1} 
\nonumber \\
& & \times \prod_{i=1}^N
\frac{1}{(1+z_i\bar z_i)^{\frac{n-1}{2}}}, 
\end{eqnarray}
where
\begin{mathletters}
\begin{eqnarray}
e^{N^{\prime}}_s & = & \prod_{k=1}^{N^{\prime}} 
\bar z_{s,k}, \\
\Psi_{s,1} & = & \prod_{k_1<k_2}^{N^{\prime}}
d(z_{s,k_1}-z_{s,k_2}),
\end{eqnarray}
\end{mathletters}
and  ${\bf \text{AN}}$ is 
the anti-symmetrizing operator on all coordinates $z_{s,k}$. 
The wave functions $\Psi=P(\phi)D\chi_n$ and
$\Psi_{\text{hard}}=\Psi_1 P(\phi -N+1)\Psi_1\chi_n$  
can be written in the following form,
\begin{mathletters}
\begin{eqnarray}
\Psi & = & {\bf \text{AN}} P \prod_{s=0}^{n-1} 
[e^{N^{\prime}}_s ]^s \Psi_{s,2}
\prod_{i<j}^N (z_i-z_j)^2
\nonumber \\
& & \times \prod_{i=1}^N
\frac{1}{(1+z_i\bar z_i)^{\frac{\phi}{2}+n-1}},  \\
\Psi_{\text{hard}} & = & \Psi_1
{\bf \text{SY}} P(\phi -N+1) 
\prod_{s} [ e^{N^{\prime}}_s ]^s \Psi_{s,2} \nonumber \\
& & \times \prod_{i<j}^N (z_i-z_j)\prod_{i=1}^N\frac{1}{(1+z_i
\bar z_i)^{\frac{\phi-N+1}{2}+n-1}},
\end{eqnarray}
\end{mathletters}
where ${\bf \text{SY}}$ is the symmetrizing operator 
on the electron coordinates, and
\begin{equation}
\Psi_{s,2}  =  \prod_{k_1<k_2}^{N^{\prime}}
(z_{s,k_1}-z_{s,k_2}).
\end{equation}
Before doing the anti-symmetrizing 
or the symmetrizing in the formulas above, it appears
that there are $n$ different groups of  electrons
and there are correlations between different groups. 
The generic terms before doing the projection, 
for example, in the formula of $\Psi$, are
\begin{equation}
\frac{\bar z_{s,k}^s z_{s,k}^l}{(1+z_{s,k}
\bar z_{s,k})^{\frac{\phi}{2}+n-1}}.
\end{equation}
It will be projected to
\begin{equation}
\frac{(\phi+1)!l!(\phi+n-1-l)!}{(\phi+n)!(l-s)!(\phi -l+s)!}
\frac{z_{s,k}^{l-s}}{(1+z_{s,k}\bar z_{s,k})^{\frac{\phi}{2}}}.
\end{equation}
As $\case{(\phi+1)!}{(\phi+n)!}$ is a constant and is not
dependent on $s, l$, we can discard it
in the process of the projection.
Thus $P$ will act in the following way (discarding
constant $\case{(\phi+1)!}{(\phi+n)!}$), 
\begin{eqnarray}
P  \frac{\bar z_{s,k}^sF(z_{s,k})}{(1+z_{s,k}
\bar z_{s,k})^{\frac{\phi}{2}+n-1}}
& = & \frac{1}{(1+z_{s,k}
\bar z_{s,k})^{\frac{\phi}{2}}}
\frac{1}{(\phi -z_{s,k}\partial_{z_{s,k}})!} \nonumber \\
& & \times 
\partial_{z_{s,k}}^s  (\phi +n-1 -z_{s,k}\partial_{z_{s,k}})!
\nonumber \\
& & \times   F(z_{s,k}).
\end{eqnarray}  
For example, by applying this formula to $\Psi =PD\chi_2$, 
the wave function is then given by 
\begin{eqnarray}
\Psi & = & {\bf \text{AN}} P \prod_{i=1}^N\frac{1}{(1+z_i
\bar z_i)^{\frac{\phi}{2}}}
\prod_{k=1}^{N/2}(\phi+1-z_{0,k}\partial_{z_{0,k}})
\nonumber \\ 
& &  \times \partial_{z_{1,k}} \prod_{s=0}^{1} (\Psi_{s,2})^3
\prod_{k_1=1}^{N/2}\prod_{k_2=1}^{N/2}(z_{0,k_1}-z_{1,k_2})^2.
\end{eqnarray}  
The trial wave function $\Psi_c$
for filling $1-\nu$ is related to
the trial wave function $\Psi$ 
at filling $\nu$ by charge conjugation,
\begin{equation}
\Psi_c=
\int\prod_i^M 
\frac{dz_{N+i}d\bar z_{N+i}}{(1+z_{N+i}\bar z_{N+i})^2}
\bar \Psi(z_{N+i}\ldots z_{N+M}) \Psi_1(z_1\ldots z_{N+M})
\label{chargec}
\end{equation}
where $M$ is the number of particles in the state $\Psi$, 
$N$ is the number of electrons in $\Psi_c$, 
$N+M=\phi+1$, and $\bar \Psi$ is the complex conjugate 
of $\Psi$. Note again, If  we use
$PD\chi_n$ as the trial wave function $\Psi$,  
the projection operator $P$ can be dropped in Eq.\ (\ref{chargec}).
However if one uses $\Psi_{\text{hard}}= \Psi_1 P(\phi-M+1)\Psi_1 \chi_n$
in Eq.\ (\ref{chargec}),  
then the operator $P$ can {\bf not} be drooped
in Eq.\ (\ref{chargec}). 

One can also act $D$ on $\Psi$ and we will get another
trial wave function of the FQH state
at filling $\frac{1}{2+\nu}$, where $\nu$ is the filling of the state
$\Psi$. Repeatedly acting $D$ and $C$  on $PD\chi_n$, 
we can get the trial wave functions at all  observable fillings.

\section{The Overlaps Between Hierarchical wave functions and
CF wave functions}
\label{fileoverl}

We perform the calculations of the wave functions
symbolically by using Maple.
The overlaps between the hierarchical wave function
and the CF wave functions are calculated, 
Some overlaps between the wave functions 
with or without the hard core construction  are also calculated.
The formula of the trial wave functions for
the FQHE in the previous sections need to be normalized
before we calculate the overlaps.
Table\ \ref{table2} lists some overlaps 
at some fillings for a small number of electrons.
$E\Psi$ means a new state formed by the condensation of
quasielectrons of parent state $\Psi$, and
$H\Psi$ means a new state formed by the condensation of
quasiholes of parent state $\Psi$. 
In all cases listed in the table,
$p_i$ is equal to $2$  for $i>1$ in the constructions of
the hierarchical states. The wave functions which involve 
$D, P, C$ operations are the CF wave functions.

When $N=3, 4, \, \phi=6$, there is only one rotational
invariant state, which must also be the ground state. 
This explains why some of overlaps 
in the table are equal to one exactly. 

Because of limited CPU we are allowed to use, we are only able to
calculate some hierarchical wave functions up to $6$ electrons,
and some CF wave functions up to $10$ electrons.  
The detailed calculations can be found 
in Ref.\ \onlinecite{carmem}. In the future, we will calculate the
wave functions  with more numbers of electrons.

From the calculations, we conclude that,  
the hierarchical wave functions and the CF wave functions
are  all most the same in the case of a small number of electrons.

\section{Conclusions}

In this paper, we present a detailed discussion about
the calculation of  the trial wave functions on the sphere.
The projective coordinates are used in performing 
the calculations.  A self-contained derivation of
the LLs on the sphere (or any surfaces with 
a constant curvature) using geometrical 
method is also given in the paper.
The many-body wave function in the LLL are studied 
and classified  in the angular momentum bases. 
We also simplify the formulas for the hierarchical wave functions 
and the CF wave functions. 

There are many interesting things which we want to study
in the future.  We shall use theories of polynomials 
to study those  wave functions.\cite{itzykson,Mac}
It would be very interesting if we can obtain 
the polynomials explicitly for  the wave functions at 
an arbitrary number of electrons.

There is a mapping  between a  trial wave function in the FQHE 
and   a wave function in  an one-dimension  
space.\cite{azuma} Because of the existence
of the mapping, one may apply the method  used to 
study the Calogero model to study the trial wave functions
in the FQHE,  and then it may be possible to calculate
some physical quantities from the trial wave functions
at an arbitrary  number of electrons.

\section{Acknowledgements}

DL thanks  Prof. P. Sodano for the discussions and
encouragements,   and Prof. R. Iengo for the discussions
concerning Landau levels on curved surfaces.
The work of DL is supported by  INFN of Italy,
and CLB  is supported by CNPQ. 
DL also thanks Prof. J. Helayel  
for hospitality  and CNPQ of Brazil for
financial support during his 
staying at CBPF in Rio de Janeiro,
where the work started.  We also thank LNCC in Rio de Janeiro
for allowing  us to use the computer facilities.

\appendix

\section*{Landau Levels on Compact Closed Surfaces}

In this appendix, we will study the LLs on general compact closed
surfaces, and work out the  LLs on the sphere as an example.

If the magnetic field  and the curvature are constant,
the spectrum, the wave functions 
and the degeneracy of Landau levels (LLs)
can be obtained by using a very 
simple geometric argument.\cite{iengoli}
A self-contained presentation of the idea
based on Ref.\ \onlinecite{iengoli} will be found in this appendix,
and some examples will be included.

In the case that the surface is a plane, a sphere, or a torus, 
the spectrum and eigenfunctions  of the LLs
can be exactly solved.\cite{mono} For example,  the LLs on a sphere 
with a Dirac-monopole on the origin,
were  solved  by Dirac long time ago. The problem in the case of 
the surface being an open up-half hyperbolic plane
with a constant negative curvature was solved completely
where there exist a  discreet spectrum  (this is
the spectrum of  the LLs)  in the low-energy sector  and  a 
continue spectrum in the high-energy sector.\cite{mono}

Ref.\ \onlinecite{iengoli}  studied the problem of 
the LLs on the compact closed 
Riemann surfaces with Poincar{\' e}  metric,
and obtained  the discrete low-energy  eigenvalues (or LLs),  
their multiplicity and  wave functions. 
Previous to Ref.\ \onlinecite{iengoli},  
similar problem also was studied 
and  discrete low-energy  eigenvalues
and their multiplicity was obtained 
by using the results from the mathematical 
literature, for example by using Selberg trace formula 
(see the references  quoted in Ref.\ \onlinecite{iengoli}).

Why the problem of the LLs  
in all those  surfaces mentioned above
can be solved completely? 
By closely following  the 
observation  in  Ref.\ \onlinecite{iengoli},
it is quite clear that the method developed in
Ref.\ \onlinecite{iengoli}  can be easily generalized 
to the case of any {\it constant}  curvature surface 
with a {\it constant} magnetic field applied on the surface
(the surface can be a compact and closed surface, or
an open surface, for example, an open up-half hyperbolic plane), 
and thus the problem of the LLs can be solved exactly in such cases.

We will show that, if the curvature and magnetic field are constant,
we can get many informations about  the spectrum and the degeneracy
of the LLs  without solving the wave functions of the LLs
explicitly by using a simple geometric argument, \cite{iengoli} 
even though the surface can be a very complicated one.
If the magnetic field is constant, 
the  wave functions  of the ground states
turn out to be a holomorphic line bundle defined on the surface.
If the curvature of the surfaces is constant too,
for the high LL,  the wave functions of the LLs
are obtained  by repeatedly applying  covariant 
derivatives on some holomorphic line bundles
(which will be specified later).  The spectrum 
is obtained without solving 
the wave functions explicitly and the degeneracy of the LLs
can be obtained by the Riemann-Roch theorem.
If the sections of some holomorphic line bundles can be obtained, 
the  wave functions of the LLs can be obtained 
explicitly.

We use  two simple examples to demonstrate how to use
this geometric approach to solve the LLs.
The examples are the LLs on the sphere and
the open up-half hyperbolic plane.

\subsection{Ground States}

We will show here that, when the magnetic field is
constant, the ground states satisfies a first-order
holomorphic (or anti-holomorphic) differential equation
and the ground states belong to the sections of 
a holomorphic line bundle.

We consider a  particle on a  
surface interacting with a  magnetic field. 
In complex coordinates,  the metric is 
$ds^2=g_{z\bar z}dzd{\bar z}$ 
and  the volume form is $dv=[ig_{z\bar z} / 2]dz\wedge d{\bar z}
= g_{z\bar z}dx\wedge dy$.  
The natural definition of the  constant magnetic 
field to the high genus Riemann surface  is
\begin{equation}
F=Bdv=(\partial_zA_{\bar z}-
\partial_{\bar z}A_z)dz\wedge d{\bar z},
\end{equation}
where   $B$ is a constant.  Thus  we have 
$ \partial_zA_{\bar z}-\partial_{\bar z}A_z =ig_{z\bar z}B / 2$.
If the surface is  closed, the magnetic 
field is then called  ``monopole" field
and subjected to the Dirac quantization condition.
The flux $\phi$ ($\phi$ must be an integer)
is given by $2\pi \phi = \int F =BV$,
where $V$ is the area of the surface and
we assume here $B>0$ ($\phi >0$) for simplicity.
The Hamiltonian of a particle on the surface is
given by the following equation,
\begin{eqnarray} 
H & = & \frac{1}{2m \sqrt{g}}
(P_{\mu}-A_{\mu})g^{\mu \nu}\sqrt{g}(P_{\nu}-A_{\nu})
\nonumber \\
& = & \frac{g^{z\bar z}}{m}
[(P_z-A_z)(P_{\bar z}-A_{\bar z})
+ (P_{\bar z}-A_{\bar z})(P_z-A_z)]  \nonumber \\
& = & \frac{2g^{z\bar z}}{m}
(P_z-A_z)(P_{\bar z}-A_{\bar z})+\frac{B}{2m}, 
\label{hamil}
\end{eqnarray}
where $g^{z\bar z}=1/g_{z\bar z}\, $,   
$P_{z}=-i\partial_z \,$,  $P_{\bar z}=-i\partial_{\bar z}\, $,
$\partial_z =(\partial_x -i\partial_y)/2 \,$,
and  $\partial_{\bar z}=(\partial_x +i\partial_y)/2 $.
We define the inner product between two wave functions as
$<\psi_1 | \psi_2 >=\int dv {\bar \psi_1 }\times \psi_2$.

Define $H^{\prime}=\case{2g^{z\bar z}}{m}
(P_z-A_z)(P_{\bar z}-A_{\bar z})$. 
$H^{\prime}$ is a positive definite hermitian operator 
because $<\psi | H^{\prime} |\psi > \, \geq 0$ for any $\psi$. 
If  $H^{\prime} \psi =0$, $\psi$  must satisfy 
$(P_{\bar z}-A_{\bar z})\psi =0$. 
The solutions of this equation
are the ground states of the Hamiltonian $H$ or $H^{\prime}$. 
In the case of closed compact surface,
the existence of the solutions of this equation 
is guaranteed by the Riemann-Roch theorem.\cite{griff,griff1}
The solutions belong to the sections of the 
holomorphic line bundle with the connection 
given by the gauge field. 
The Riemann-Roch theorem tells us that 
\begin{equation}
h^0(L)-h^1(L)=deg(L)-h+1, 
\label{riroch}
\end{equation}
where $h$ is the genus of the surface,
$h^0(L)$ is the dimension of the sections of
the holomorphic line bundle
or the degeneracy of the ground states of the Hamiltonian $H$,   
$h^1(L)$ is the dimension of the holomorphic differential 
$(L^{-1}\times K)$, where $K$ is the canonical bundle, 
and $deg(L)$ is the degree of the line bundle,  which is equal to
the first Chern number of the gauge field, or the magnetic flux
out of the surface, $\phi$.  
When $deg(L) >2h-2$,  $h^1(L)$ is equal to zero,\cite{griff}
thus $h^0(L)=\phi-h+1$.  One finds that $h^0(L)$ indeed 
gives the right degeneracy of the ground states in the case of  
a particle on a sphere or a  torus interacting 
with a magnetic-monopole field.

In the case of non-compact surfaces, 
for example an infinite plane or an  up-half hyperbolic plane,
the flux out of the surfaces are infinite, 
and the degeneracy is infinite too.
The degeneracy of the LLs turns out be infinite. 
Thus Eq.\ (\ref{riroch})
also gives  correctly the degeneracy,  
as when  the flux is infinite, 
the equation implies that $h^0(L)$ becomes infinite.   
When the surface has a boundary, for example
a disc, one would expect  that Eq.\ (\ref{riroch})
is replaced by  a new index relation 
given by the  boundary index theory.  Note that,
when the flux is much bigger than one, 
the  degeneracy of the ground states are approximately 
equal  to the flux $\phi$ out of the surface.

\subsection{Higher Landau Levels }

We study the higher LLs in the 
case of  the curvature of the surface 
being constant. When the curvature is constant,
$g^{z\bar z}\partial \bar \partial ln g_{z\bar z}=C$,
a Liouville-like integrable equation. 
For the flat surface,  $C=0$, as in the case 
of a plan or a torus,   
the spectrum and the  wave functions 
of the LLs can be completely solved.
When the surface is flat, the higher 
LLs are obtained by applying successively
a first order  differential 
operator to the  states in the LLL.
Now we shall generalize such construction of
the LLs in the case of a flat surface to
the case of a curved surface. 

We consider  here the {\it closed} and {\it curved}  surface 
with constant (non-zero) curvatures.  
It is easy to generalize to the case of 
an open surface  with a constant curvature
and we will demonstrate it in 
an example in the end of the appendix. 
When $C$ is not equal to zero,  one has 
$g_{z\bar z}=(1/C) \partial \bar \partial ln g_{z\bar z}$.
As  the magnetic field is constant, 
we can fix the gauge field as
$A_z=-i B^{\prime}\partial (\ln g_{z\bar z}) /2$,
and the magnetic field  $F$ is equal 
to $Bdv$, where $B=2B^{\prime}C$. 
For example, in the case of the Poincar{\' e} metric,
$ds^2=y^{-2} (dx^2+dy^2)\, $,   $g_{z\bar z}=y^{-2}\,$, and 
$C=1/2$, thus $B=B^{\prime}$. 
For  a closed  surface, by Gauss theorem, the flux $\phi$
out of the surface is equal to $\phi=B(h-1)/c=2B^{\prime}(h-1)$.
$B^{\prime}$ must be a rational number as $\phi$ is an integer.
For the negative curvature closed surface, 
according to  Gauss theorem,  
we should have $h \geq 2$. On the other hand,
for the positive constant curvature surface, 
$h$ must be equal to zero, 
and thus the surface is topologically equivalent to a sphere. 
Without losing any generalities,  
we assume in the following discussions
that  $B$ is a positive number. For a negative $B$, 
the wave functions are the complex
conjugate of the wave functions in the case of a positive $B$.

For any eigenfunctions of the Hamiltonian,  they satisfy
\begin{equation} 
H \psi = E \psi . 
\end{equation}
If the domain of $\tilde z$ intersects non-trivially the
domain of $z$, $ g_{z\bar z}dz d\bar z$ is
invariant under coordinate changes, or
\begin{equation}
g_{z\bar z}dz d\bar z =
g_{\tilde z \tilde {\bar z} }d{\tilde z} 
d\tilde {\bar z}
\end{equation}
on the intersection of the domains of $z$ and $\tilde z$. Define 
\begin{equation}
D=\partial - (B^{\prime}/2) \partial \ln g_{z\bar z}\, ,
\bar D= \bar \partial + (B^{\prime}/2) \bar \partial 
\ln g_{z\bar z}.
\end{equation}
$D$ and $\bar D$ are transformed as
\begin{equation}
{\tilde D}=(d z/d 
{\tilde z})U^{-1} D U \, \, ,
\tilde {\bar D}=(d\bar z/d \tilde
{\bar z}) U^{-1} \bar D U  
\end{equation}
where $U(z, \tilde z)= (\case{dz}{d \tilde z})^{-B^{\prime}/2}
(\case{d \bar z}{d  \tilde {\bar  z}})^{B^{\prime}/2}$.

We take $m=2$ in Eq.\ (\ref{hamil}) for the simplicity.
The Hamiltonian can be written in the following form,
\begin{equation} 
H = -g^{z\bar z} D \bar D +(B/ 4). 
\label{hamil1}
\end{equation}
Thus the Hamiltonian in  the domain $z$ 
is transformed  to the Hamiltonian 
in  the domain $\tilde z$  as
\begin{equation}
\tilde H =U^{-1} H U, 
\end{equation} 
and the wave function is transformed as
\begin{equation}
\tilde \psi =U^{-1} \psi.
\end{equation} 
Therefore $\psi (dz)^{B^{\prime}/2}(d\bar z)^{-B^{\prime}/2}$
is invariant under the transformation, and it implies 
that $\psi$ is a differential form of type
$T^{\bar B^{\prime}/2}_{B^{\prime}/2}$, where we use the following
notation:  if $F(z,\bar z)(dz)^{X}(d\bar z)^{Y}$ 
is invariant under the transformation, then $F(z,\bar z)$  is  
a differential form of type $T^{-\bar Y}_{X}$.

The ground states are given by the solutions of the
equation $ \bar D \psi =0$.  
When the curvature is negative, 
$C, \phi,  B^{\prime}$ are  positive numbers.
If $\phi > 2h-2$, or  $B^{\prime}>1$,  
then the number of the solutions is $\phi - h+1$
according to the previous discussions.
%(remind that $h>1$ in this case).
For smaller $\phi$, some discussions can be found in 
Ref.\ \onlinecite{iengoli}. In the case of compact and closed 
Riemann surfaces with the Poincar{\' e} metric, 
the  wave functions in LLL were  constructed by calculating
the determinant of holomorphic sections of some bundle.\cite{iengoli}
 
When the curvature is positive, 
$C$ and $B^{\prime}$ are  negative numbers, and $h=0$ as shown 
in the previous discussions.  Now $\phi$ 
is equal to  $\phi =-2B^{\prime}$. 
The LLL states are again given by the solutions of the
equation $ \bar D \psi =0$.  
As  $|\phi | > 2h-2=-2 $ ($h=0$ in this case), 
the number of the solutions is equal to $|\phi |-h +1=|\phi | +1$.

To  obtain the spectrum and wave functions of the higher LLs,
we introduce the covariant derivative, \cite{iengoli} 
$\nabla_z$, and its Hermitian conjugate  
$(\nabla_z)^{\dagger}=-\nabla^z$,
\begin{mathletters}
\begin{eqnarray}
& \nabla_z & : T^{l}_k \to 
T^{l}_{k+1} \, \, ,  \nabla_z =g^k\partial g^{-k},   \\
& (\nabla_z)^{\dagger} &: T^{l}_k \to T^{l}_{k-1} \, \, ,   
(\nabla_z)^{\dagger} =-
g^{-l-1} \bar \partial g^{l},
\end{eqnarray}
\end{mathletters}
where we call $g=g_{z\bar z}$ for short.
Note that $D$  is the covariant operator $\nabla_z$  acting on
$T^{ \bar B^{\prime} /2}_{B^{\prime}/2}$, and
$\bar D = g \nabla^z$ where $\nabla^z$ 
acts on $T^{ \bar B^{\prime} /2}_{B^{\prime}/2}$.
The Hamiltonian can be written by using the covariant operators,   
\begin{equation}
H-B/4 =-\nabla_z \nabla^z .
\end{equation} 
One can verify the commutation relation,
\begin{equation}
[\nabla^z \, \, \nabla_z]T^m_n
= -(m+n)C. 
\label{crel}
\end{equation}
Assume that $\psi_1$ is a state in the higher LLs 
and an eigenfunction of $H$
with eigenvalue $E_1$,
then $\psi_1 =-\case{1}{\epsilon_1}\nabla_z \nabla^z \psi_1$, 
where $\epsilon_1 =E_1-B/4 > 0$. 
Therefore one can write $ \psi_1 =\nabla_z \Phi (1)$, where
$\Phi (1)$ is a differential form of type 
$T^{\bar B^{\prime}/2}_{B^{\prime}/2-1}$. 
More explicitly, we have
\begin{equation}
\psi_1 = (\partial - (B^{\prime}/2-1)\partial ln g)\Phi (1) .
\end{equation}
Using the relation 
$\partial \bar \partial ln g=gC$, one can  show that,
\begin{equation}
-\nabla_z \nabla^z \psi_1 =(B^{\prime}-1)C\psi_1 + 
\nabla_z [-\nabla_z \nabla^z \Phi (1)].
\end{equation}
We first discuss  the case of 
a negative curvature surfaces.
If $B^{\prime}\geq 1$, one can show that  
$<\psi_1|\nabla_z [-\nabla_z \nabla^z \Phi (1)]>\geq 0$.
Thus one can conclude  that the states of 
the lowest excited level are obtained,  if 
there exist $\Phi$ such that $\nabla_z \nabla^z \Phi (1)=0$.
$\nabla_z \nabla^z \Phi (1)=0$ leads to $\bar D\Phi (1)= 0$. 
The solution of $\bar D\Phi (1)= 0$ is
$\Phi (1) =g^{-B^{\prime}/2}
\tilde \Phi (1)$ with $\bar \partial \tilde \Phi (1) =0$, 
where $\tilde \Phi (1) $ is of the form $T_{B^{\prime}-1}$.
By the Riemann-Roch theorem,   there exist 
solutions of the equation $\bar \partial \tilde \Phi (1) =0$
for $B{\prime}\geq 1$, and  the number of the solutions or
the degeneracy of this Landau level  is the  dimension of the
sections of the holomorphic bundle $T_{B^{\prime}-1}$,
which is equal to $(2B^{\prime}-3)(h-1)$ if $B^{\prime}>2$.
The energy of this LL or the  lowest excited states is 
\begin{equation}
E_1={3\over 2}B^{\prime}C -C.
\end{equation}
If $B^{\prime}<1$, there is only 
the  $zero'th$ ``Landau level" or the LLL
(there exists a continue spectrum in the high-energy sector,
and the states in the continue spectrum are not called
as the states in the LLs).

We can generalize the above discussion to higher LLs.
The wave function of the $k'th$ LL is given by
\begin{eqnarray}
\psi_k & = & (\nabla_z)^{k}\Phi(k) \nonumber \\
&=& (\partial-(B^{\prime}/2-1)\partial \ln g)
(\partial-(B^{\prime}/2-2)\partial \ln g)  \nonumber \\
&  & \times \cdots
(\partial-(B^{\prime}/2-k)\partial \ln g)\Phi (k),  
\label{LLs}
\end{eqnarray}
with $\Phi (k)=g^{-\case{B^{\prime}}{2}} \tilde \Phi (k) $ and
$\bar \partial \tilde \Phi (k)=0$. 
$\tilde \Phi (k)$
is a differential form of the type $T_{B^{\prime}-k}$. 
Notice that this construction 
generalizes the standard construction for the harmonic oscillator.
The difference for the constructions
of the high LLs between the case of the flat surfaces 
and the case of the curved surfaces is clear now.
In the case of the surface being a plan or a torus,
the high LLs are obtained by applying successively a 
first order differential operator to the ground states.
However  the situation is different when the surface
is curved.  $\Phi (k)$ for $k\not= 0$ is 
{\it not} the ground state of the Hamiltonian $H$.

Using Eq.\ (\ref{crel}),
we calculate the eigenvalue of the corresponding 
wave function $\psi_k$ ,  and it is equal to
\begin{equation}
E_k=CB^{\prime}(k+{1 \over 2})-{k(k+1)C \over 2}. 
\label{kthen}
\end{equation}
The  degeneracy of the $k'th$ LL
is given by the dimension of the sections of the holomorphic
bundle of   the type $T_{B^{\prime}-k}$, which is equal to
$(2B^{\prime}-2k-1)(h-1)$ when $B^{\prime}-k>1$.  
Because the dimension of
$T_n$ is zero when $n$ is negative,
$k$ must not be greater than $B^{\prime}$. 
Hence there is only a {\it finite}  number of  ``Landau levels".

When $B^{\prime}$ is an integer, $k$ can take value
from  $0$ to $B^{\prime}$. When $k=B^{\prime}$, 
the corresponding $\tilde \Phi (k) $
is the differential form of the type $T_0$.
$T_0$ is a constant function on the surface
and the  degeneracy of this LL is equal to one.
For the twisted  boundary conditions, 
which would physically correspond to the presence of 
some magnetic flux through the handles.
There does not exist a non-zero constant function
which satisfies the twisted boundary condition,
thus the dimension of $T_0$ is zero,  and 
the  degeneracy of this LL is equal to zero or
there does not exist this LL 
($B^{\prime}-th$  LL).
When $k=B^{\prime}-1$, the degeneracy of this LL
is the dimension of the canonical bundle
$T_1$, which is equal to $h$
for the non-twisted  boundary condition and is equal to $h-1$
for the twisted  boundary condition (this result can be 
obtained by the Riemann-Roch theorem).
$B$ could be also an half-integer.
Then $k$ can take value from $0$ to $B^{\prime}-(1/2)$. 
When $k=B^{\prime}-(1/2)$,  the degeneracy of this LL
is the dimension of the spin bundle
$T_{1/2}$. The dimension of the 
holomorphic sections of the spin bundle
generically is zero for the even-spin structures
and one for the odd ones (or for twisted ones).
It is possible that  $B^{\prime}$ is fractional assuming
that $2B^{\prime} (h-1)$ is an integer, and 
$k$ can take value from $0$ to $[B^{\prime}]$
where $[B^{\prime}]$ is the bigger integer which is smaller
than $B^{\prime}$. When $k=[B^{\prime}]$,
the degeneracy of this LL
is the dimension of  bundle $T_{B^{\prime}-[B^{\prime}]}$. 
$B^{\prime}-[B^{\prime}]$ is a fractional number
between $0$ and $1$ and  the discussions  of such case
can be found in Ref.\ \onlinecite{iengoli}.
Beyond those LLs, little is known about the continue spectrum
in the case of  the complicated negative curvature surfaces.

To normalize $\psi_k$, we calculate 
the inner product $<\psi_k | \psi_k >$.
By using Eq.\ (\ref{crel}). It is given 
by the following equation,
\begin{eqnarray}
<\psi_k | \psi_k > & = & <\nabla_z^{k}\Phi (k)
|\nabla_z^{k}\Phi (k)> \nonumber \\
&=& <\Phi (k)| (\nabla_z^{k})^{\dagger}\nabla_z^{k}\Phi (k)>
\nonumber \\
&=  & <\Phi (k) |\Phi (k)>C^k 2^{-k}k! \nonumber \\
& & \times \prod_{i=1}^{k} (2B^{\prime}-k-i),
\end{eqnarray}
where the inner product  $<\Phi (k) | \Phi (k) >$ is defined as
\begin{equation}
<\Phi (k) | \Phi (k) >=\int dv g^k
{\bar \Phi (k) }\times \Phi (k).
\label{innerphi}
\end{equation}
The definition of  the inner product 
between two $\Phi (k)$
given in Eq.\ (\ref{innerphi}) is 
quite natural  because $\Phi (k)$  
is a differential form of the type 
$T^{\bar B^{\prime}/2}_{(B^{\prime}/2) -k}$.

If $\Phi (k)$ is normalized to one, then 
\begin{equation}
\psi_k \over 
[C^k k!2^{-k}\prod_{i=1}^{k}
(2B^{\prime}-k-i)]^{1/2} 
\end{equation}
is also normalized to  one.

Now we come to  the case of a closed 
surface with a positive curvature,
which is a little bit different from the case 
of a surface with a negative curvature.
Now we have only $h=0$ according to the previous discussion.
The wave function $\psi$ is a differential form of type
$T^{\bar B^{\prime}/2}_{B^{\prime}/2}$
with $B^{\prime}$ being a  negative number.
In the formula $-\nabla_z \nabla^z \psi_1 =
(B^{\prime}-1)C\psi_1 + \nabla_z [-\nabla_z \nabla^z \Phi (1) ]$,  
one can show that  
$<\psi_1|\nabla_z (-\nabla_z \nabla^z \Phi )>\geq 0$
for any negative $B^{\prime}$.
By using Riemann-Roch theorem, one finds that
there {\it always}  exists $\Phi (1)$ such that 
$\nabla_z \nabla^z \Phi (1)=0$, which leads to
$\bar D\Phi= 0$.  Therefor for any  $B^{\prime}$,
there exists a higher LL. One can repeat the argument
to obtain the states in the higher LLs and  
obtains the {\it full} spectrum and wave functions. 

The wave functions  of the states in the $k'th$ LL 
are  again given by Eq.\ (\ref{LLs}), 
with $\tilde \Phi (k)=g^{B^{\prime}/2} \Phi (k)$ and
$\bar \partial \tilde \Phi (k)=0$. 
$\tilde \Phi (k)$ is a differential 
form of the type $T_{B^{\prime}-k}$.
The degeneracy of the $k'th$ LL is equal to
the dimension of the holomorphic line bundle
$T_{B^{\prime}-k}$, which is equal to
$2(B^{\prime}-k)(h-1)-h+1=-2(B^{\prime}-k)+1$  
as $h$ is equal to zero.

The energy is again given by  Eq.\ (\ref{kthen}). However,
a higher LL  has a {\it higher} degeneracy  
and the number of the LLs is  {\it infinite} in such case. 
Instead, in the case of a surface with 
a negative curvature, 
a higher LL has a {\it smaller} degeneracy.
and the number of the LLs is {\it finite}.
From Eq.\ (\ref{kthen}), one  notices that, 
in the case of a positive
curvature surface, the energy gap in the neighboring LLs
{\it increases} when  the level increases, 
and in the case of a negative
curvature surface, the energy gap in the neighboring LLs
{\it decreases} when  the level increases.

It is easy to generalize the 
above discussions to non-compact surfaces, and 
we will work out an example in the following discussion.

\subsection{Examples}

\subsubsection{Upper half hyperbolic surface}

We consider that the surface is a upper half hyperbolic surface 
(also see   Comtet and  Dunne 
in Ref.\ \onlinecite{mono}). 
In the projective coordinates. 
the metric $g$ is written as $\case{1}{(1-z\bar z)^2}$, 
where $|z|\leq 1\,$,  The other quantities are,
$C=2\,$,   $B=4B^{\prime}\, $,  and
$A_z=-i\case{B^{\prime}{\bar z}}{(1-z\bar z)^2}$.
The  wave functions are given by
Eq.\ (\ref{LLs}). As the wave functions of the LLs
shall be normalizable (opposite to the wave function 
of a state inside the continue spectrum), 
$<\psi_k | \psi_k >$ shall be normalizable.
A normalizable $<\psi_k | \psi_k >$ is 
equivalent to a normalizable $<\Phi (k) | \Phi (k) >$.
A normalizable $<\Phi (k) | \Phi (k) >$  
leads the condition $B^{\prime}-(1/2)>k \geq 0$. 
$\Phi (k)$ is given by function $g^{-B^{\prime}/2}z^l$
where $l$ is a non-negative integer. 
Thus the degeneracy is infinite for every LL.
This is consistent with the Riemann-Roch theorem
as the flux out of the surface is infinite.
Finally, the energy is given by  Eq.\ (\ref{kthen}).

\subsubsection{Sphere}

Another example is that the surface is a sphere.
In the projective  coordinates.
the metric $g$ is written  as $g=\case{1}{(1+z\bar z)^2}$.
The other quantities are, 
$A_z= i\case{B^{\prime}{\bar z}}{(1+z{\bar z})^2}\,$, 
$C=-2 \, $, and $B=2B^{\prime}C =-4B^{\prime}$.
Thus the flux $\phi =-2B^{\prime}$  
is a non-negative integer according to the Dirac
quantization condition (note we always assume $B>0$ in this paper).
The  wave functions are again given by
Eq.\ (\ref{LLs}), the energies are given by Eq.\ (\ref{kthen}), and
$\Phi (k)$ is given by $g^{\phi /4}z^l$.
The normalizable condition leads to $l=0, 1, \cdots , 2k+\phi$.
Thus the degeneracy of the $k'th$ LL
is equal to $2k+\phi +1$.   The degeneracy can be also obtained 
by the Riemann-Roch theorem and the result 
is consistent with the result obtained by
requiring the wave functions being normalizable.
In this way, we obtain the full spectrum 
and all  wave functions on the sphere.

The wave functions at  the $n'th$ Landau level 
($n=0$ is the lowest Landau level) are given by
Eq.\ (\ref{allLLs}).

From previous discussions, we can 
easily find the inner product
$<\psi_{n,l}|\psi_{n,l}>$ is equal to
$\pi \case{l!(\phi+2n-l)!}{(\phi +2n+1)(\phi +n)!}$.
The inner product is as previously defined, 
$<\psi_1 | \psi_2 >=\int dv {\bar \psi_1 }\times \psi_2$,
where $dv =\int \case{dxdy}{(1+z \bar z)^2}$.

However inside the paper,   The definition of 
the inner product is {\it different} from
the definition in the appendix.
The inner product in the paper is {\it defined}  as
\begin{equation} 
<\psi_1 | \psi_2 >=\int \case{dzd\bar z}{(1+z \bar z)^2}
{\bar \psi_1 }\times \psi_2.
\end{equation}
As $ dzd\bar z =2 dxdy$,  thus
$<\psi_{n,l}|\psi_{n,l}>$ is given by the following formula, 
\begin{equation}
<\psi_{n,l}|\psi_{n,l}>=2\pi 
\frac{l!(\phi+2n-l)!}{(\phi +2n+1)(\phi +n)!}.
\end{equation}
This formula is used in the paper.

\mediumtext
\begin{table}
\caption{In this table, we list the number of rotational
invariant states at various $\nu$ and a small number of
electrons. $N_t$ is the dimension 
of the total Hilbert space (in the LLL)
and $N_r$ is the number of the rotational invariant  states.}
\label{table1}
\begin{tabular}{c|c|c|c|c|c|c} 
$\nu$ & $N$ & {\em $N_t$} & {\em $N_r$} & $\phi$ (formula)  &
$\phi$ & $ \frac{N\phi}{2}$  \\
\tableline
$\frac{2}{5}$ & 4 & 5 & 1 & $\frac{5}{2}N-4$ & 6 & 12 \\
              & 6 & 58& 3 & & 11 & 33 \\
              & 8 & 910 & 8 & & 16 & 64 \\ \hline
$\frac{2}{7}$ & 4 & 43 & 2 & $\frac{7}{2}N-2$ & 12 & 24 \\
              & 6 & 1.242 & 10 & & 19 & 57 \\
              & 8 & 46.029 & 80 & & 26 & 104 \\ \hline
$\frac{2}{9}$ & 4 & 43 & 2 & $\frac{9}{2}N-6$ & 12 & 24\\
              & 6 & 2.137 & 13 & & 21 & 63\\
              & 8 & 139.143 & 164 & & 30 & 120\\ \hline
$\frac{2}{11}$  & 4 & 150 & 3 & $\frac{11}{2}N-4$ & 18 & 36\\
                & 6 & 11.963 & 29 & & 29 & 87\\
                & 8 & 1.229.093 & 702 & & 40 & 160\\ \hline
$\frac{2}{13}$  & 4 & 150 & 3 & $\frac{13}{2}N-8$ & 18 & 36\\
                & 6 & 17.002 & 34 & & 31 & 93\\
                & 8 & 2.502.617& 1.137 & & 44 & 176\\ \hline
$\frac{3}{7}$  & 9 & 910 & 8 & $\frac{7}{3}N-5$ & 16 & 27 \\ \hline
$\frac{3}{11}$  & 6 & 2.137 & 13 & $\frac{11}{3}N-1$ & 21 & 63\\
                & 9 & 610.358 & 506 & & 32 & 144\\ \hline
$\frac{3}{17}$  & 6 & 17.002 & 34 & $\frac{17}{3}N-3$ & 31 & 93 \\ \hline
$\frac{5}{17}$ & 4 & 33 & 2 & $\frac{17}{5}N-\frac{13}{5}$ & 11 & 22 \\
               & 9 & 184.717 & 217 & & 28 & 126 \\ 
\end{tabular}
\end{table}

\widetext
\begin{table}
\caption{The  overlaps between the hierarchical wave functions
and the  CF wave functions at some fillings 
for a small number of electrons.}
\label{table2}
\begin{tabular}{c|c|c|c|c|c} 
$\nu$ & $N$ & \multicolumn{4}{c} {\em overlap} \\ 
\tableline
$\frac{2}{5}$ &  & $<PD \chi_2|E\Psi_3>$  
&  $<\Psi_1P\Psi_1 \chi_2|E\Psi_3>$ 
& $<PD \chi_2|\Psi_1E\Psi_1^2>$ & 
$<\Psi_1P\Psi_1 \chi_2|\Psi_1E\Psi_1^2>$   \\ 
& 6 & .9993234149 & .9993615971 & .9998331523 & .9999456457 \\
& 4 & 1 & 1 & 1 & 1\\ 
\end{tabular}
\vspace{1cm}
\begin{tabular}{c|c|c|c||c|c|c} 
$\nu$ & $N$ & \multicolumn{2}{c}{\em overlap} 
& $\nu$ & $N$ & {\em overlap} \\ 
\tableline
$\frac{2}{5}$ &  & $<E\Psi_3|
\Psi_1E\Psi_1^2>$ & $<PD \chi_2
|\Psi_1P\Psi_1 \chi_2>$  
& $\frac{2}{7}$ & & $<H\Psi_3|DC\Psi_3>$ \\
& 6 & .9996479001 & .9999288987 & & 6 & .9993762574 \\
& 4 & 1 & 1 & & 4 & 1  \\ 
\end{tabular}
\vspace{1cm}
\begin{tabular}{c|c|c|c|c} 
$\nu$ & $N$ & \multicolumn{3}{c}{\em overlap} \\ 
\tableline
$\frac{2}{9}$ & & $<E\Psi_5|
DPD\chi_2>$ & $<E\Psi_5|
PD^2\chi_2>$ & $<DPD\chi_2|PD^2\chi_2 >$ \\
& 4 & .9999614869 & .9999614869 &  1 \\
\end{tabular}
\vspace{1cm}
\begin{tabular}{c|c|c||c|c|c}
$\nu$ & $N$ & {\em overlap} & $\nu$ & $N$ & {\em overlap}  \\ 
\tableline
$\frac{3}{11}$ & & $<EH\Psi_3|DCPD\chi_2>$ &
$\frac{2}{11}$ & & $<H\Psi_5|D^2C\Psi_3>$ \\ 
& 6 & .9996522383 & & 4 & 1  \\ 
\end{tabular}
\vspace{1cm}
\begin{tabular}{c|c|c||c|c|c}
$\nu$ & $N$ & {\em overlap} & $\nu$ & $N$ & {\em overlap} \\ \hline
$\frac{2}{13}$ & & $<E\Psi_7|DPD^2\chi_2>$ &
$\frac{5}{17}$ & & $<HH\Psi_3|DCDC\Psi_3>$ \\
& 4 & .9999218859 & & 4 & .9999999997 \\ 
\end{tabular}
\end{table}

\begin{references}
\bibitem[*]{byline0}E-mail: souzbati@cbpfsu1.cat.cbpf.br
\bibitem[\dag]{byline}E-mail: lidp@pg.infn.it
\bibitem {Laughlin2}R.B. Laughlin, Phys. Rev. Lett. 
{\bf 50},  1395 (1983).
\bibitem {PrangeGirvin}R.E. Prange and S.M. Girvin, 
{\it The Quantum Hall Effect} (Springer-Verlag, New York, 1990);
R. Laughlin, in {\it Fractional Statistics and Anyon Superconductivity},
edited by F. Wilczek (World Scientific, Singapore, 1990). 
\bibitem {Haldane605}F.D.M. Haldane, 
Phys. Rev. Lett. {\bf 51}, 605 (1983).
\bibitem {Halperin1}B.I. Halperin, 
Phys. Rev. Lett. {\bf 52}, 1583 (1984).
\bibitem{frast}D. Arovas, R. Schrieffer and F. Wilczek, 
Phys. Rev. Lett {\bf 53}, 722  (1984).
\bibitem{other}S. Girvin, Phys. Rev. B {\bf 29},  6012 (1984);
A.H. MacDonald and D.B. Murry, {\it ibid.} {\bf 32},  2707 (1985);
M.P.A. Fisher and D.H. Lee, Phys. Rev. Lett. {\bf 63}, 903 (1989).
\bibitem{Read}N. Read, Phys. Rev. Lett. {\bf 65}, 1502 (1990).
\bibitem{wenblok}B. Blok and X.G. Wen, 
Phys. Rev. B{\bf 42}, 8133 (1990);
 {\bf 42}, 8145 (1990); {\bf 43}, 8337 (1991).
\bibitem{greiter}M. Greiter, Phys. Lett. B {\bf 48}, 3336 (1994). 
\bibitem{jian}J. Yang, Phys. Rev. B {\bf 50}, 11196  (1994).
\bibitem {Jain3}J.K. Jain, Phys. Rev. Lett. {\bf 63}, 199 (1989);
Phys. Rev. B {\bf 41}, 7653 (1990); Adv. Phys. {\bf 41}, 105  (1992). 
\bibitem{wenr}X.G. Wen, 
%``Topological Orders and Edge Excitations in FQH States", 
preprint, cond-mat/9506066, and references therein.
\bibitem{torus}F.D.M. Haldane and E.H. Rezayi, Phys. Rev. B {\bf 33},
3844 (1986); D. Li, Int. J. Mod. Phys. B {\bf 7},  2779 (1993);
{\bf 7}, 2655 (1993);
E. Keski-Vakkuri and X.G. Wen, {\it ibid.} {\bf 7}, 4227 (1993).
\bibitem{ha}F. Lesage, V. Pasquier and D. Serban, Nucl. Phys. B 
{\bf 435}, 585 (1995); Z.N.C. Ha,  {\it ibid.} {\bf 435},  604 (1995).
\bibitem{wuyang}T.T. Wu and C.N. Yang, Nucl. Phys. B {\bf 107},
365 (1976).
\bibitem {Fano} G. Fano, F. Ortolani and E. Colombo, 
Phys. Rev. B {\bf 34},  2670 (1986).
\bibitem {LiEsf}D. Li, Nucl. Phys. B(FS) {\bf 396},  411 (1993);
R. Iengo e K. Lechner, Phys. Reports {\bf 213},  179 (1992).
\bibitem {iengoli} R. Iengo and D. Li, Nucl. Phys. B(FS) 
{\bf 413}, 735 (1994); and references therein.
\bibitem {GirvinJach} S.M. Girvin and T. Jach, Phys. Rev. B 
{\bf 29},  5617 (1984).
\bibitem{rezayi}B.I. Halperin, P.A. Lee, and N. Read,
Phys. Rev. B {\bf 47},  7312 (1994); E. Reazayi and N. Read,
Phys. Rev. Lett. {\bf 72},  900 (1994); N. Read, preprint, 
cond-mat/9501090, and references therein.
\bibitem {Andrews} G.E. Andrews, {\it The Theory of Partitions} 
(Addison-Wesley Publishing Company, Massachusetts, 1976).
\bibitem{itzykson1}C. Itzykson, 
%``Interacting  Electrons  In  A Strong Magnetic Field",
In  {\it Quantum Field Theory and Quantum Statistics}, Vol. 1, 
edited by I. A. Batalin, {\it et al.}
(Adam Hilger, Bristol, 1987); 
%p. 415-448;
B. Derrida and J.P. Nadal, J. Phys. Lett. (Paris) {\bf 45},  701 (1984).
\bibitem{hard}J.K. Jain and X.G. Wu, preprint, cond-mat/9312091.
\bibitem{arovas}D. Arovas, %Topics in Fractional Statistics, 
in {\it Geometric Phase in Physics}, 
edited by A. Shapere and F. Wilczek
(World Scientific, Singapore, 1989)
\bibitem{carmem}Carmem Lucia de Souza Batista, Master thesis, CBPF,
Rio de Janeiro, 1996.
\bibitem{itzykson} P. Di Francesco, M. Gaudin, C. Itzykson and
F. Lesage,  Int. J. Mod. Phys. A {\bf 9}, 4257 (1994);
G. Dunne, Int. J. Mod. Phys. B {\bf 7}, 4783 (1993).
\bibitem{Mac}I.G. MacDonald, {\it Symmetric 
Functions and Hall Polynomials} 
(Claredon Press, Oxford, 1979).
\bibitem{azuma}H. Azuma and S. Ito, Phys. Lett. B {\bf 331}, 107 
(1994); P.K. Panigrahi and M. Sivakumar, 
Phys. Rev. B {\bf 52}, 13742 (1995). 
\bibitem{mono}P. Dirac, Proc. R. Soc. London A {\bf 133}, 60 (1931);
I. Tamm, Z. Phys. {\bf 71},  141 (1931);
M. Fierz, Helv. Phys. Acta. {\bf 17},   27 (1944);
Ref.\ \onlinecite{wuyang};
A. Comtet, Ann. Phys. {\bf 173}, 185 (1987), and references therein;
G.V. Dunne, Ann. Phys. {\bf 215},  233 (1992), and references therein;
J. Zak, Phys. Rev. {\bf 134},  A1607  (1994).
\bibitem{griff}E. Arbarello, M. Cornalba, P.A. Griffiths
and J. Harris, {\it  Geometry of Algebraic Curves},  Vol. 1
(Springer-Verlag, New York, Berlin, Heidelberg, Tokyo, 1985).
\bibitem{griff1}H.M. Farkas and I. Kra, {\it Riemann Surfaces}
(Springer-Verlag, Berlin,  1980).
\end{references}
\end{document}